\documentclass [10pt]{iopart}
\usepackage{iopams}
\newcommand{\p}{$p$}
\newcommand{\pnorm}[1]{\left\vert#1\right\vert_p}
\newcommand{\Qp}{\mathbb{Q}_{p}}

\begin{document}

\title[\p-adic description of characteristic relaxation in complex
systems]
{\p-adic description of characteristic relaxation in complex systems}%
\author{V A Avetisov, A Kh Bikulov, V A Osipov}%
\address{N N Semenov Institute of Chemical Physics, RAS,
ul.Kossygina 4, 119991 Moscow, Russia} \ead{avetisov@chph.ras.ru}
\begin{abstract}
This work is  a further development of an approach
to  the description of relaxation processes in complex
systems  on  the basis of the  \p-adic analysis.
We show that three types of relaxation fitted into
the Kohlrausch-Williams-Watts law, the power decay law, or the
logarithmic decay law, are   similar random processes.
Inherently, these processes  are ultrametric and are described by the \p-adic
master equation. The  physical meaning  of this equation is explained
in terms of  a random walk constrained by a
hierarchical energy landscape. We also discuss relations  between the
relaxation kinetics and the energy landscapes.
\end{abstract}
\submitto{JPA}  \maketitle

\section{Introduction.}

One of the distinctive features  of complex systems (such as
 glasses, macromolecules, and proteins)  is an
anomalously slow non-exponential relaxation observed in a very wide range
of time scales ---  from that of short-time molecular vibrations to
days and months. Relaxation kinetics in complex systems is often
fitted into  three empirical laws, namely, the stretched
exponential curve (the Kohlrausch-Williams-Watts
law),~${\sim\exp\left[- \left({t\left/\tau\right.}\right)^{\alpha
}\right]}$, ${0 <\alpha<1}$, the power
decay law,~${\sim\left(t\left/\tau\right.\right)^{ - \alpha }}$, ${0
<\alpha}$, and the logarithmic decay law,~${\sim \alpha \left[ln
\left(t\left/\tau\right.\right)\right]^{ - 1}}$,
${1<\alpha}$~\cite{Ngai, Brawer, Matsuoka}. These types of
relaxation are  characteristic of complex systems, just as
exponential relaxation  is typical  for  gases and liquids.

Although  formulation of a microscopic theory of anomalous
relaxation in complex systems  remains an issue of the day, there
exist important concepts regarding the general relaxation behavior
of complex systems.  The starting point of these concepts is that
the states of  a micro-component of a system are strongly coupled
to configurational rearrangements of the local environment,  and
it is important that these rearrangements cover a wide  range of
scales --  from microscopic to mesoscopic. In order to describe
configurational dynamics of this kind, the concept of energy
landscapes is applied~{\cite{Stillinger}-\cite{Frauenfelder2}}.
According to this concept,  configurational rearrangements are
described by a random walk in the space of configurational states,
and this process, in its turn, is determined by the energy
landscape of the system.

There is  common understanding  of the fact that a comprehensive
description of energy landscapes of complex systems is hardly
possible for several  reasons. First of all, due to an
uncertainty while selecting local potentials of interaction
between the system's components. Another reason is the large
number of metastable configurations, which occur due to multiple
quenched constraints and the random nature  of the local arrangement. The
number of metastable configurations  grows exponentially
with respect to  the size of the system.  Thus,  energy landscapes of mesoscopic
systems (these systems are  of the main interest)
form  multidimensional surfaces with an extremely large number of
local minima. It is not quite clear how to describe those surfaces
analytically  and, accordingly, how to describe the
configurational dynamics constrained by such landscapes. In this connection,
computer simulation of  model systems remains the main field of
activity. These studies show that  energy landscapes of complex
systems may  vary significantly. However, the characteristic
types of relaxation obviously suggest that such landscapes have some common features.
 These features might turn out to be as
fundamental for the dynamics of complex systems as the space
symmetry is for  crystal
structures.

One of the nontrivial assumptions, which initially appeared in the
theory of spin glasses~\cite{Parisi} - \cite{Mezard},  was that the
energy landscapes of complex systems have a  hierarchical structure.
In recent years, this conjecture was confirmed by the results
of numerous computer simulation studies (see, for
example,~\cite{Becker, Wales}). It turned out that  energy
landscapes of complex systems represent hierarchically nested
basins of local minima,   viz.,  large basins consist of smaller basins
of minima, the latter consist of even smaller ones, etc. The
basins of minima are separated from one another by a hierarchy of
activation barriers, i.e.,  the larger basins are separated by
higher activation barriers, while the smaller basins within the
larger ones  are separated by smaller activation barriers,
respectively.

The existence   of the basins of local minima allows one to
describe  configurational dynamics in terms of the basin-to-basin
kinetics~\cite{Becker}.  In this connection, it is assumed that
transitions between the basins belonging to the same level of the
basin's hierarchy correspond to configurational rearrangements of
a certain scale. Within such basins, a quasi-equilibrium
distribution is established during relatively short (for a given
scale) period of time, while transitions between the basins
determine long-time configurational rearrangements. The hierarchy
of nested basins corresponds to the hierarchy of scales of
configurational rearrangements.

As indicated in the early works on the spin-glass theory, the
hierarchy of nested basins of local minima is similar to the
hierarchy of nested ultrametric
spheres~\cite{Parisi}-\cite{Mezard}. This idea can be found in
some  publications about  anomalous relaxation in glass-like
systems~{\cite{Ogielski}-\cite{Kohler}}. However, these
investigations  give no answer to the question whether
ultrametricity can be directly used  for the description of the
dynamics of complex systems.

Recent studies~\cite{ABK, ABKO} show  that there is an
approach which may  lead to a noticeable progress in this direction. This
approach is based on the  \p-adic analysis natural for
 ultrametric spaces.   An introduction to the \p-adic analysis can
be found, for instance, in~\cite{VVZ}. Some  applications
of the \p-adic analysis in theoretical physics and
theoretical biology are  described
in~{\cite{Vladimirov}-\cite{Parisi1}}.

The \p-adic master equation describing a random walk constrained
by the hierarchical energy landscapes of a certain type (the
so-called ultrametric diffusion) was constructed in~\cite{ABK,
ABKO}. In these works, we  describe  a general method of finding
its solutions and examine  some  Cauchy problems pertaining to
conformational dynamics and reactions in proteins, and we also
find some solutions of these problems expressed by analytic
formulas.  It is an interesting fact that the \p-adic
pseudo-differential equation introduced earlier as a \p-adic
analogue of  the diffusion equation (see~\cite{VVZ}) coincides
with the particular  \p-adic  master equation
 obtained in~\cite{ABK, ABKO}.

Regarding \p-adic pseudo-differential equations and stochastic
processes in non-Archimedean spaces, see ~\cite{VVZ},
\cite{Albeverio}-\cite{Kochubei2}.

The present paper is aimed at a further development of
the approach suggested in~\cite{ABK, ABKO}  for the description of
configurational dynamics of complex systems.     We show  that the
Kohlrausch-Williams-Watts law, the power decay law, and the
logarithmic decay law  can be  described by  \p-adic master
equations of the same form, and therefore,  the relaxation laws characteristic of
 complex systems  actually reflect the same type of ultrametric random processes.

The paper is structured as follows. In  Section 2, we  introduce  a
general form of the \p-adic master equation describing  random
walk constrained by a hierarchical energy landscape, and we also
 define  a \p-adic model of relaxation. In Section 3, we consider  three
versions of such models  and establish their correspondence with
 three types of  relaxation in complex systems.

\section{Random walk in ultrametric space}

The general  \p-adic master equation describing a Markovian
process of random walk in ultrametric space can be written as
follows:
\begin{equation}
\label{eq1} \frac{\partial f(x,t)}{\partial t} = \int_{\Qp}
{\left[ {w(x\vert y)f(y,t) - w(y\vert x)f(x,t)} \right]d\mu (y)},
\end{equation}

\noindent where~$x$ is a \p-adic number, $d\mu (x)$~is the Haar
measure on the field of \p-adic numbers~$\Qp$, and~$t$ is time,
${t \in \mathbb{R}_{ + }}$. Equation~(\ref{eq1}) is the usual
balance equation for  transitions between  system states.
The only peculiarity is that the space of states is described by
the \p-adic numbers, i.e. this space is ultrametric. The function
${f(x,t) :\; \Qp \times \mathbb {R}_+ \mapsto \mathbb {R}_+}$ is a
probability density distribution: the  integral~$\int_{B}f(x,t)d\mu
(x)$ is the probability of finding  the system in a domain~$B
\subseteq \Qp$ at the instant~$t$. The function ${w\left(x\vert y
\right):\;\Qp \times \Qp \mapsto \mathbb{R}_+}$ is the probability
of the transition from the state~$y$ to the state~$x$ per unit time.

The transition from a state~$y$ to a state~$x$ can be perceived as
overcoming  the energy barrier separating these states. The structure
of the relation between  the transition probability ${w\left( {x\vert y}
\right)}$ and the parameters of the energy barrier depends, in general,
 on the mechanism of configurational rearrangements,
transition pathways, their distribution, etc. Finding this
expression on the basis of the microscopic theory is a nontrivial problem
beyond the scope of this article.  For the sake of definiteness,
we will use  the standard  Arrhenius relation
\begin{equation}\label{eq2}
w\left( {x\vert y} \right)\sim A\left( T \right)\exp
\left\{ { - \frac{U\left( {x\vert y} \right)}{kT}} \right\}
\end{equation}
\noindent where~$U\left( {x\vert y} \right)$ is the height of the
(effective) activation barrier for the transition from the state~$y$ to
the state~$x$, $k$~is the  Boltzmann constant, and $T$~is temperature.
Studies of thermal behavior of anomalous relaxation in
complex systems show that this expression is quite acceptable
for a fairly wide range of temperatures.

Formula  (\ref{eq2}) establishes a  relation between the structure
of the energy landscape $U\left( {x\vert y} \right)$ and the
transition function $w\left( {x\vert y} \right)$. In particular,
the condition ${w\left( {x\vert y} \right)=w\left( {y\vert x}
\right)}$ corresponds to a degenerate energy landscape. An
important example of a degenerate landscape is a regular
hierarchical landscape. In this case, on each hierarchical level,
the basins of states  split up into the same number of smaller
basins; moreover,   the activation barriers between the basins of
the same  hierarchical level have equal values. Graphically, this
hierarchical landscape is represented  as a regular tree with the
same branching rules for all branching nodes. As  shown
in~\cite{ABK, ABKO}, the activation barriers $U\left( {x\vert y}
\right)$ separating the states~$x$ and~$y$ depend, in this case,
only on  ultrametric distances between the states, ${\pnorm{x -
y}= p^\gamma}$, ${\gamma \in \mathbb{Z}}$, and therefore,  regular
energy landscapes are completely described by  functions of the
form $U(\pnorm {x- y})$. Thus, in the case of random walk
constrained by a regular hierarchical landscape, the master
equation~(\ref{eq1}) takes the form
\begin{equation}
\label{eq3} \frac{\partial f\left( {x,t} \right)}{\partial t} =
\int_{\Qp } {W(\pnorm{x - y})\left[ {f(y,t) - f(x,t)}
\right]d\mu(y)}
\end{equation}
\noindent where
$$
W( \pnorm{x - y}) = \frac{A\left( T \right)}{\pnorm{x - y}}\exp
\left\{-\frac{U\left( \pnorm{x - y} \right)}{kT}\right\}\;.
$$
The \p-adic master equation~(\ref{eq3}) admits analytic  solutions
for a fairly wide class of  functions $W(\pnorm{x - y})$. The
technique of finding its solutions is described, for instance,
in~\cite{ABK, ABKO}.

It may seem that regular hierarchical landscapes represent a very
idealistic model of real landscapes and have little relation to
the latter. Indeed, regular hierarchical landscapes reflect no
details of the energy landscapes of complex systems other than
their  hierarchical structure. Oddly enough, this is sufficient
for the description of the characteristic types of relaxation.

On the basis of general physical considerations, one can identify
three essentially different types of regular hierarchical
landscapes:

``Logarithmic landscapes'' are characterized by activation
barriers having a slow (logarithmic) growth with respect to the
number~$\gamma$ of the hierarchical level, ${U(\gamma)\sim \ln
\gamma}$. This type of landscapes may be associated with
``flexible'' systems having a wide range of available scales for
configurational rearrangements.

Contrary to logarithmic landscapes, ``exponential landscapes'' are
characterized by a rapid  (exponential) growth of activation
barriers, ${U(\gamma)\sim e^{\gamma}}$. Landscapes of this type
represent ``stiff''  systems with a small set of distinct scales
available for configurational rearrangements.

Finally, there are ``linear landscapes'' with linear growth of
activation barriers, ${U(\gamma)\sim \gamma}$. These landscapes
may be associated with systems of an ``intermediate'' type.

In all these  cases, we  consider the Cauchy problem for
the respective master equation~(\ref{eq3}) with the initial
conditions:
\begin{equation}
\label{eq4} f(x,0) = \Omega \left( \pnorm{x} \right) = \cases{
 1& $\pnorm{x} \le 1$\\0& $\pnorm{x} > 1$\;.}
\end{equation}
A  relaxation process will be understood as the  evolution of
population  in the domain of the initial
distribution~${\pnorm{x}\le 1}$:
\begin{equation}\label{eq5}
S(t) = \int\limits_{\pnorm{x}\le 1} {f(x,t)d\mu(x)}\;.
\end{equation}
In each of the above cases, the analytical estimation of
relaxation function~$S(t)$ is being found. It would enable the
relation of~$S(t)$ to three empirical laws for characteristic
relaxation.

\section{Characteristic  relaxation types}
The solution of the Cauchy problem for  equation~(\ref{eq3}) with
the initial condition~(\ref{eq4}) has the form
\begin{equation}\label{eq6}
\fl f(x,t)=\sum_{n=0}^\infty e^{\tilde W(p^{-n})
t}\left[(1-p^{-1})p^{-n}
\Omega(p^{-n}\pnorm{x})-p^{-n-1}\delta(\pnorm{x}-p^{1+n})\right]
\end{equation}
\noindent where
$$
\delta(\pnorm{x}-p^{\gamma})=\cases{ 1&$ \pnorm{x}=p^{\gamma}$\\
0& $\pnorm{x}\neq p^{\gamma}$\;. }
$$
Indeed, applying the \p-adic Fourier transformation to~(\ref{eq3})
and taking into account the initial condition~(\ref{eq4}), we get
$$
f(x,t)=\int_{\mathbb{Z}_p}e^{\tilde{W}(\pnorm{k})t}\chi(-kx)d\mu(k)
$$
\noindent where
\begin{equation}\label{eq7}
\tilde{W}(\pnorm{k})=\int_{\Qp}W(\pnorm{x})(\chi(kx)-1)d\mu(x)\;.
\end{equation}
Calculating  the integral and using the substitution ${\pnorm{k}=p^{-n}}$,
we obtain
$$
\tilde{W}(p^{-n})=-p^{n+1}\left[(1-p^{-1})\sum_{m=0}^\infty p^m
W(p^{m+n+1})+{p^{-1}}W(p^{n+1})\right]
$$
\noindent and therefore, (\ref{eq6}) is the desired solution.

In order to find $S(t)$, the solution~(\ref{eq6}) should be
substituted into~(\ref{eq5}).  Calculating the  corresponding
integral, we find that
\begin{equation}\label{eq8}
S(t)=(p-1)\sum_{n=1}^\infty p^{-n}e^{-\Lambda_n t}\;.
\end{equation}
\noindent Here, for the sake of convenience, we have introduced
the positive function
\begin{equation}\label{eq9}
\Lambda_n =p^n\left[(1-p^{-1})\sum_{m=0}^\infty p^m
W(p^{m+n})+p^{-1}W(p^n)\right].
\end{equation}

Next, we obtain estimates of  $S(t)$  for  the above three regular
hierarchical landscapes.  The transition
probability $W$ is defined here by (\ref{eq2}). The frequency factor is taken
equal to the unity, and the temperature dependence is characterized  by the
parameter~${\alpha= T_{0}\left/T\right.}$.

\subsection{Logarithmic landscape.}
In view of (2),
the transition
probability function for the logarithmic landscape can be represented as
$$
W(\pnorm{x-y})=\frac{1}{\pnorm{x-y}}\;\frac{1}{\ln^\alpha(1+\pnorm{x-y})}
\;,\quad\alpha>1\;.
$$
\noindent In this case,
\begin{equation}\label{eq10}
\Lambda_n=(1-p^{-1})\sum_{m=0}^\infty
\ln^{-\alpha}(1+p^{m+n})+p^{-1}\ln^{-\alpha}(1+p^{n})\;.
\end{equation}
Then, the following estimates hold:
$$
(1-p^{-1})\frac{\ln^{-\alpha}p}{(\alpha-1)}\frac{1}{(n+1)^{\alpha-1}}<\Lambda_n<
\frac{\alpha\ln^{-\alpha}p}{(\alpha-1)}\frac{1}{n^{\alpha-1}}\;.
$$
Using these inequalities, we can estimate  the
relaxation function~$S(t)$ as follows:
\begin{eqnarray*}
\fl p\;\exp\left\{-\frac{2\alpha}{\alpha-1}\left(\frac{t}{\ln
p}\right)^{\frac{1}{\alpha}}\right\}\\
<S(t)<p^{2}\left(\exp\left\{-\frac{1}{\alpha-1}\left(\frac{t}{\ln
p}\right)^{\frac{1}{\alpha}}\right\}+
\exp\left\{-\left(\frac{t}{\ln
p}\right)^{\frac{1}{\alpha}}\right\}\right)\;.
\end{eqnarray*}
\noindent Thus, \textit{the long-time behavior of relaxation
kinetics constrained by logarithmic landscapes is fitted into
the Kohlrausch-Williams-Watts law.}

It is interesting to note, that as $\alpha\to1$, the upper
bound for $S(t)$  approaches an  exponent. For
${\alpha \le 1}$, the series~(\ref{eq10}) is divergent. This
fact  may be interpreted as a transition to  exponential
relaxation  when a critical temperature $T_{0}$ is attained. This kind of
critical behavior of ultrametric diffusion constrained by
logarithmic landscape was previously predicted in~\cite{Ogielski}
on the basis of some  qualitative considerations.

\subsection{Linear landscape.} In the case   of a linear
landscape, the transition probability function has the form:
$$
W(\pnorm{x-y})=\frac{1}{\pnorm{x-y}^{\alpha+1}}\;.
$$
Estimates
for the probability density distribution $ f(x,y)$
 in the case of a linear landscape were
obtained in~\cite{ABKO}.
Calculations for $S(t)$ similar to those of~\cite{ABKO} yield the estimate
$$
\frac{1}{p}\Gamma\left(\frac{1}{\alpha}+1\right)\left(-\Gamma_p(-\alpha)t\right)
^{-\frac{1}{\alpha}} <S(t)
<\Gamma\left(\frac{1}{\alpha}+1\right)\left(-\Gamma_p(-\alpha)t\right)
^{-\frac{1}{\alpha}}
$$
\noindent where $\Gamma(\alpha)$ is the gamma-function and
$\Gamma_p(\alpha)$ is the \p-adic gamma-function (see~\cite{VVZ}).

Thus, \textit{the long-time behavior of relaxation kinetics
constrained by  linear landscapes is fitted into the power
decay law.}

\subsection{Exponential landscape.} In this case, the
transition probability function is
$$
W(\pnorm{x-y})=\frac{1}{\pnorm{x-y}}\;e^{-\alpha\pnorm{x-y}}\;,
\quad\alpha>0,
$$
\noindent and from (\ref{eq9}) we have
$$
\Lambda_{n}=\left(1-\frac{1}{p}\right)\sum^{\infty}_{m=0}\exp\left[-\alpha
p^{n+m} \right]+\frac{1}{p}\;\exp\left[-\alpha p^{n}\right]\;.
$$
The following estimates hold:
$$
\exp\left[-\alpha
p^{n}\right]<\Lambda_{n}<\frac{p^{\alpha}}{p^{\alpha}-1}\exp\left[-\alpha
p^{n} \right]\;.
$$
Hence,  for relaxation function $S(t)$ we get
$$
\frac{p}{2e}\;\frac{\alpha}{\ln t+\ln
\left(p^\alpha/(p^\alpha-1)\right)} <
S(t)<(p^2+p+e)\frac{\alpha}{\ln t}\;.
$$
The last estimate  shows that \textit{the long-time behavior of
relaxation kinetics constrained by  exponential landscapes is
fitted into the logarithmic decay law.}

\nosections It has been  shown  that the
Kohlrausch-Williams-Watts law, the power decay and the logarithmic
decay laws  reflect the same type of random processes. Such processes  are
ultrametric and can be described by the \p-adic master equation
formulated in~\cite{ABK, ABKO}.  It seems
  that ultrametricity is
a most general and fundamental feature of the dynamics of complex systems:
The configurational space of such systems may be
totally disconnected   due to numerous quenched constraints,
and therefore,  it is impossible to accomplish a relatively large configurational
rearrangement  step-by-step, by way of small configurational
changes.

In conclusion, we consider it necessary to make a few remarks.

A direct comparison of the function $S(t)$ with  experimentally observed
relaxation kinetics requires a certain caution. Indeed, the function $S(t)$
is the population of a region in the configurational space, whereas
 in most experiments, the focus of
observation is some probe  coupled to  configurational
rearrangements, rather than  configurational rearrangements
themselves.  A specific  form of the relation   between the probe  state
and a local configuration  may happen  to be important for the description of
 experimental data.

The above three characteristic relaxation laws  do not
exhaust the diversity of relaxation kinetics in  complex systems.
There are numerous observations of "mixed" types of kinetics, with
one type of  relaxation kinetics  observed in some
time-window and  another type  observed outside.  For
instance, it may happen  that at low temperatures the observed
kinetics is well fitted into the logarithmic decay law, and at higher
temperatures it is described  by  the power decay law. In other cases, the
power decay may transform into the Kohlrausch-Williams-Watts law.
It is not difficult to imagine  energy landscapes whose activation
barriers near the ground state are close to the exponential
landscapes, while higher activation barriers correspond  to a linear
landscape  and then transform  into a logarithmic landscape.
The above approach can be easily used
for  the description of
ultrametric diffusion constrained by such "mixed" landscapes.
 \Eref{eq3} can be  solved analytically  for a wide
class  of functions ${W(\pnorm{x-y},\alpha)}$ admitting
 the \p-adic Fourier transformation.

Finally, our considerations here are  restricted to degenerate  energy
landscapes. These  may be directly related only to some specific  complex
systems, for example, spin glasses. However,  energy landscapes
of other complex systems  are obviously non-degenerate. Local
minima in such landscapes are also clustered into a hierarchy of
basins, and as the whole, they look like a global potential
``hole'' (or ``funnel'') with extremely rugged ``walls''. The
description of random walk on such landscapes by \p-adic equations
is of great interest, particularly, in connection with the
applications of the concept of hierarchical energy landscapes to
the protein folding~\cite{Bringelson} (see also~\cite{Huang} and
references therein). Some results  regarding this problem will be
presented in our further  publications.

\ack We would like to express our profound gratitude to
Academician V.S.Vladimirov and Professor I.V.Volovich for fruitful
discussions and valuable remarks. The authors are grateful to Dr.
A.P.Zubarev and Dr. S.V.Kozyrev for their active participation in
discussions. The given work has been partially supported by
RFBR-00-15-97392, RFBR-00-15-96073, and INTAS-9900545 grants.


\end{document}